# Lightweight Security Protocol for WiSense based Wireless Sensor Network

Sahilkumar Chakraborty
Department of Electronics and Communication
RV College of Engineering
Bangalore, India

Astha Nachrani
Department of Computer Science and Engineering
RV College of Engineering
Bangalore, India

Aronee Dasgupta
Department of Telecommunication
RV College of Engineering
Bangalore, India

Pritam Gajkumar Shah, PhD
Department of Computer Science and Engineering
RV College of Engineering
Bangalore, India

## ABSTRACT
Wireless Sensor Networks have emerged as one of the leading technologies. These networks are designed to monitor crucial environmental parameters of humidity, temperature, wind speed, soil moisture content, UV index, sound, etc. and then transfer the required information to the base station. However, security remains the key challenge of such networks as critical data is being transferred. Most sensor nodes currently deployed have constraints on memory and processing power and hence operate without an efficient security protocol. Hereby a protocol which is lightweight and is secure for wireless sensor applications is proposed.

## 1. INTRODUCTION
Due to developments in analogue and digital technologies many corporations have developed low power and low cost wireless sensor nodes which are used in sensing of environmental parameters [1]. Sensor nodes process the information gathered and then transmit it to the coordinator or to other slave nodes to provide intelligence for better understanding of the environment (depending upon the network topology) [2]. Applications include perimeter monitoring, vehicle emission monitoring, defence monitoring (in reconnaissance scenarios), etc. Openness of the channel and low physical protection of the channel poses a security risk.

Current encryption standards have been designed for high performance devices and are taxing on the sensors which have low computation capabilities and have low available bandwidth. More so, security protocols, developed for ad hoc networks cannot be applied directly for sensor networks due to difference in their architecture design. Sensor networks are self-organizing, dynamic, with peer to peer data transmission capabilities sensor networks on the other hand have a well-defined topology with a central base station to which all the data is relayed. Employing traditional encryption algorithms in the sensor would cause extra bits to be processed and transmitted and due to the large processing delay incurred jitter and lag would occur throughout the network [3]. There is a paramount need for developing a robust encryption which does not required heavy cryptography processes but still provides sufficient security for the required application.

Implementing a robust yet lightweight algorithm is needed not just in the central base station but also the sensor nodes. There was a controversy regarding the implementation of AES algorithm for encryption and decryption in sensor networks utilizing the ZigBee protocol because AES needs high processing power and memory usage especially in nodes containing low amounts of available RAM [4]. Also the management and modification of the encryption keys in the sensor nodes is a major consideration as nodes are spatially distributed and remain away from direct human. The proposed protocol uses a combination of substitution using a S-Box matrix and an XOR operation with a key to provide a reasonable level of security for low performance nodes.

## 2. LITERATURE REVIEW

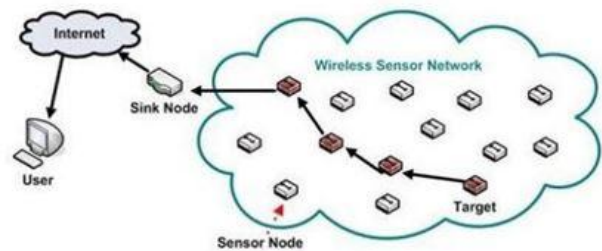

**Fig. 1. Wireless Sensor Network Architecture**

Fig.1 [5] shows a typical wireless sensor network architecture consisting of various sensor nodes connected to the sink node. The sink can be controlled by a remote terminal over the Internet. Sensor networks comprise many nodes spatially distributed over a large area. Sensor nodes are designed to work for long durations of time (possibly stretching into months) to collect, analyse and return data to base station [5].

The sensor network may have the nodes designed as passive or active. In a passive node the base station listens to the node (and the node transmits the data back to the base station) only for a fixed duration in a time interval. This reduces the battery consumption of the node as the node can run in idle or low-power mode when the base station is not listening. A duty cycle is defined for the passive node sensor network. A smaller duty cycle helps in achieving better energy efficiency in the sensor network which is desirable [6]. In an active node the base station continuously listens to the node and the node processes, analyses and returns the data gathered back to the base station.

In mobile sensor network each node needs to know each its location either through GPS or any other relative positioning algorithm. For ad-hoc fixed networks network information has to be dynamically updated in real time as sensor nodes may fail or be added to the network. The transmission protocols and software design have to be constructed keeping in mind the effects of shallow fading, number of links, communication length, mode of propagation and the minimum acceptable quality of service. Since the sensor nodes are highly susceptible to environmental parameters (like wind speed, humidity, rain, etc.) redundancy in the number of nodes have to be provided so that the network doesn't go offline.

Security protocols have to be built into the initial design of the nodes and not as a retrofit as most nodes operate in hostile conditions and it is imperative their presence remains undetected. They are limited in their available energy (around 20 – 30 joules). Since data is arriving from large number of sources the incoming data needs to be combined with the local information so that the resulting information can be passed on to the central node. Dissemination of the large amounts of data collected and processed needs the network to have to low amounts of transmission latency. Security protocols have to be designed which protects the network from spoofing and intrusion.

## 2.1 WiSense Node Architecture

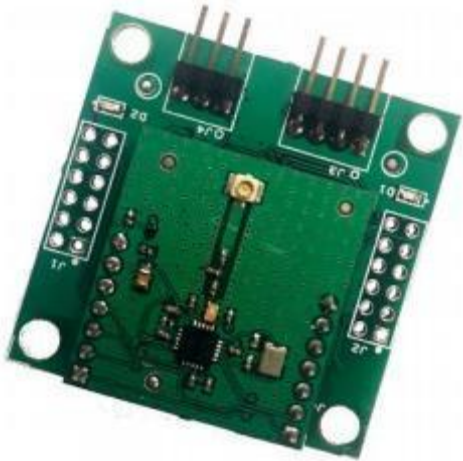

**Fig. 2. WiSense Node –courtesy WiSense [7]**

Fig. 2 shows the WiSense WSN1101L [7] based wireless mesh node. It consists of the microcontroller module (MSP430G2955) and the radio module (CC1101). The radio module has an antenna for receiving and transmitting the data. The MSP430G2955 is an ultra-low power microcontroller from TI with 56 KB of flash memory and 4 KB static ram memory. The standby current can reach as low as 1 micro ampere. Operating voltage of the microcontroller lies between 1.8 V – 3.6 V. The module has an on-chip 10-bit ADC channels. Peripheral support is provided on the module in the form of UAR, I2C and SPI. There is a SPI and GPIO interface to the radio module and UART or I2C or SPI or GPIO interface to sensors. An on-board 32KHz crystal provides the clock. The radio module (CC11011) operates below the 1 GHz frequency. It is a low cost transceiver suitable for remote sensing applications. The frequency range in which it operates lies in the 300-348 MHz, 387-464 MHz and 779-928 MHz bands [8]. On-off keying (OOK) and flexible (Amplitude-shift keying) ASK shaping are the two modulation schemes supported for RF transmission and receiving. Data rates range from 0.6 to 600 kbps. The on-board antenna is an omni directional whip antenna, having a length of 117 mm. An on-board high accuracy 26 MHz crystal provides the clock for the circuit.

## 2.2 Issues in Implementing AES in Sensor Nodes

Symmetric techniques have the problem of key agreement, issues regarding the scalability of keys and do not provide nonrepudiation [9]. AES is considered slow and requires a lot of memory for the storing the lookup tables. AES has higher mean values for energy-latency product, the energy per bit for encryption process is around 151nJ/bit for 128 bit AES encryption. [10]. Even if the AES encryption takes only 11 cycles the full program with data retrieval, encryption and calculation takes a full 704 clock according to Hodjat. The decryption process was found to take 20-30% more energy than the decryption process. Also symmetric techniques like AES, are highly architecture dependent [11].

## 2.3 Limitations in implementing AES in WiSense Nodes

The WiSense [7] node under consideration, is built on the TI MSP430G2955 a 16-bit RISC architecture based microcontroller [12]. There exist serious limitations in implementing symmetric encryption algorithms in 16-bit microcontrollers like the MSP430. Most of these algorithms are adapted for 32-bit processors, however the node under consideration is a 16-bit microcontroller. Rijndael cipher block have been optimized for 32-bit processors but these optimizations cannot be applied to the TI MSP430 due to its limited instruction set. Also these optimizations even if applied will take majority of the memory available from the node. The MSP430 has a simple but limited instruction set. The majority of register operations take up 1 cycle, but accessing the actual memory is costly and takes up to 3 cycles. Considering an example if standard 128-bit implementation of AES in TelosB note [13] had been implemented, it would take 1.994 ms for encryption and a further 2.365 ms for decryption [14].

## 3. NETWORK ARCHITECTURE
### 3.1 WiSense Network

A multihop mesh WiSense Sensor-Actuator Network (WISAN) has been considered which consists of 25 nodes with total of 72 sensors operating at 2405Mhz with a transmission power of 5 dbm. The sensor nodes collect data regarding temperature, voltage and pressure. The coordinator node has a MAC ID of 0xDEADBEEFFEEDDADD [15]. Table 1 shows the sensor information concerning the last and final update time tags. Table 2 shows the location and address of the nodes as they are spatially distributed in the environment. The node is based on the TI MSP430G2955 microcontroller. The module consists of two PCB's. One PCB hosts the microcontroller while the other hosts the CC1101 radio. The microcontroller PCB has 56 KB of flash, 4 KB of SRAM with a standby current (in LMP3) as low as 1 micro amp. It has as an optional on board serial to USB interface (to provide both power and serial connectivity). It has peripheral support in the form of SPI/I2C/UART with a 32 kHz crystal as a real-time clock [7].

### 3.2 WiSense Network

The firmware elements consist of the Reduced Function Device (RFD), Full Function Device (FFD), Coordinator and the Gateway [16]. Reduced Function Device (RFD) is essentially a mote that has sensors and firmware code for remote connectivity and local processing. The available sensor information is relayed to the FFD from the RFD. Full Function Device (FFD) is similar to a RFD, but it also helps as an intermediate relay node. A RFD sends information to a neighbouring FFD, which deals with routing that information to its neighbouring FFDs. This chain proceeds until the coordinator is reached. The PAN coordinator is the master device and the control point in the system. Various systems can exist in the same area, however, there will be one and only coordinator for every system. RFDs and FFDs partner with the coordinator when they join that specific system. The coordinator allocates every hub a PAN address that is legitimate inside the system. Messages from reduced and full function devices are sent to the gateway. Control messages are prepared by the coordinator however sensor information is generally sent to gateway the for transmission towards LAN/WAN.

## 4. IMPLEMENTATION

In the Advanced Encryption Standard (AES) with Cipher Block Chains, the outcome of encryption of the first block is used to encrypt the next. The process is continued iteratively. Such a design calls for a feedback. Each round of encryption uses the following processes viz. *SubBytes, ShiftRows, MixColumns* and *AddRoundKey* [17]. There are in total ten rounds with the last round skipping the *MixColumns* step. At the receiver end a mathematical inverse of each step is done, however, the order is reversed. Block size of 128 bits is considered. Other than the *SubBytes* step all the other three steps are linear. In the only nonlinear step (*SubBytes*) the Galois inverse is calculated of an 8-bit number which itself lies in the Galois Field GF ($2^8$). Despite being faster the look up table method suffers from one critical disadvantage: it occupies 256 bytes of storage not considering the memory needed for addressing and storing the results. [18]. Each 16 bytes in the block needs to processed separately and in parallel for the substation. However, for practical applications this would cause to 16 copies of the S-Box to be present for a single round itself and for a full pipeline implementation, 160 copies of the S-Box table for the entire encryption process [17]. A total of 40960 bytes of memory would be needed for the *SubBytes* step alone, which is expensive in sensor nodes.

### 4.1 Proposed Protocol

**Fig. 3. AES S-Box [19]**

Fig. 3 [19] shows an AES S-Box. Every sensor node in the network will have an inbuilt 16*16 S-Box [20] which would occupy a total of 2 kilobits of flash memory. Now consider two nodes that wish to communicate with each other for sending and receiving humidity, temperature, wind speed, soil moisture content, UV index or sound values etc., this value to be sent is first substituted with the corresponding element in the 16*16 S-Box by the method of substitution. The substituted value is then EX-ORed with a key, which can be - a) hour/minute/second value in the timestamp b) the node name c) last two bits of its MAC address. At the receiver end the received data is again first EX-ORed with the key then mapping is done with the help of the S-Box to arrive at the unencrypted value. Only the first 2 digits of the MAC address, timestamp, or the node name, is considered as the key to save on memory. If timestamp was to be used as the key then an extra flag should be transmitted to specify whether the hours, mins or seconds is considered as the time stamp value. The above proposed security protocol consumes the least memory of little more 2 Kilobits (along with some memory needed for the runtime code).

### 4.2 Example

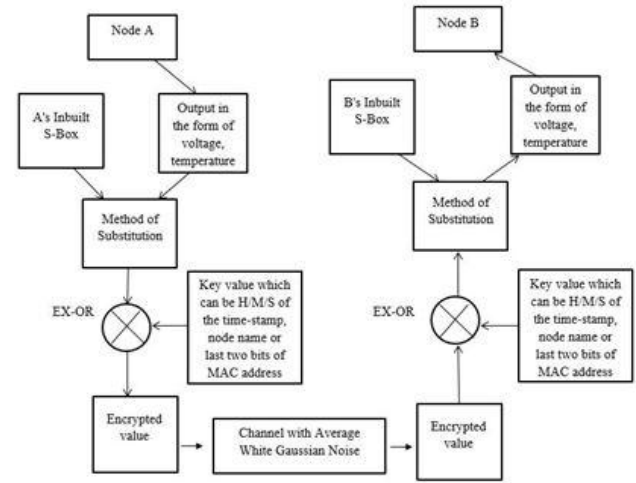

**Fig. 4. High level Block Diagram of the proposed encryption and decryption process**

Fig. 4 shows the high-level implementation of the proposed encryption and decryption process with a block diagram. The time-stamp, node name, or the MAC address can also be used as the key for the Ex-OR operation. If the time-stamp is to be considered as the key then, an additional flag of 2-bit length is needed to specify whether hours, minutes or seconds is being used as the time-stamp value. This not only creates an extra overhead, but is also susceptible to noise in the channel. Any interference in the channel may cause bit flipping and will cause the wrong time stamp to be used at the receiver for the Ex-OR operation in the decryption process.

Suppose 8-bit data: 0xD5 is to be transmitted from the node 08 with a MAC address of 0x0008000800080008 (from Table 2) with the corresponding encryption and decryption process. The data first undergoes the substitution process with the S-Box, so the substituted value after the *SubBytes* operation is 03 in hexadecimal notation. The substituted value is then EX-ORed with the last two digits of the MAC address of that node (which is 0x08). The resulting encrypted value to be transmitted is 0x0B which is 8 bit in size. The data undergo packetization. The packet also contains the physical address of the source node which is the combination of IP address and the MAC address of the sending node. At the receiver node deframing is done and the 8-bit encrypted data is EX-ORed with the last digits with the MAC address. In this case, the result is 0x03. This result is now mapped to the inverse S-Box table to arrive at the decrypted value of 0xD5.

## 5. CONCLUSION

With the widespread adoption of wireless sensor networks to monitor and gauge environmental parameters it has become necessary to come up with a simple yet robust security protocol for hardware constraint sensor nodes. The protocol developed was tested on the MSP430 based WiSense nodes and was found to be satisfactory in terms of performance and security. The combination of *SubBytes* and EX-OR provides a strong security against intrusion by unauthorized sources.

However, further optimizations can be done to provide an optimal, secure communication infrastructure for wireless sensor platforms.

**Table 1. Sensor Information with update times [16]**

| Name | Id | Last Update | Next Update |
|---|---|---|---|
| 0002 | 1 | 2015-10-18 04:04:57 | - |
| 0013 | 1 | 2015-10-18 02:03:13 | 2015-10-18 04:03:13 |
| 0019 | 4 | 2015-10-18 01:38:33 | 2015-10-18 04:38:33 |
| 0014 | 1 | 2015-10-18 01:16:00 | 2015-10-18 04:16:00 |
| 0005 | 3 | 2015-10-17 23:49:14 | 2015-10-18 00:49:14 |
| 0004 | 2 | 2015-10-17 22:38:40 | 2015-10-18 00:38:40 |
| 000F | 2 | 2015-10-17 18:03:57 | 2015-10-17 18:08:57 |
| 0013 | 5 | 2015-10-17 17:03:04 | 2015-10-17 20:03:04 |
| 0006 | 2 | 2015-10-17 16:32:14 | 2015-10-17 17:32:14 |
| 0010 | 3 | 2015-10-17 15:48:41 | 2015-10-17 15:49:11 |
| 0010 | 2 | 2015-10-17 15:48:40 | 2015-10-17 15:50:40 |
| 0010 | 1 | 2015-10-17 15:48:40 | 2015-10-17 17:48:40 |
| 0004 | 3 | 2015-10-17 15:48:10 | 2015-10-17 16:48:10 |
| 0008 | 3 | 2015-10-17 15:48:04 | 2015-10-17 17:48:04 |
| 0018 | 2 | 2015-10-17 15:11:41 | 2015-10-17 18:11:41 |
| 0002 | 2 | 2015-10-17 12:50:35 | - |
| 0004 | 1 | 2015-10-17 12:43:50 | - |
| 0008 | 1 | 2015-10-17 09:25:46 | - |
| 000D | 4 | 2015-10-17 09:04:03 | 2015-10-17 12:04:03 |
| 0017 | 4 | 2015-10-17 08:59:58 | 2015-10-17 09:59:58 |
| 0017 | 3 | 2015-10-17 08:59:58 | 2015-10-17 09:04:58 |
| 0017 | 1 | 2015-10-17 08:59:57 | 2015-10-17 10:59:57 |
| 0008 | 2 | 2015-10-17 08:53:28 | - |
| 0011 | 3 | 2015-10-17 07:30:49 | 2015-10-17 10:30:49 |
| 000D | 2 | 2015-10-17 07:30:45 | - |
| 0006 | 4 | 2015-10-17 06:42:23 | - |
| 0012 | 1 | 2015-10-17 05:48:30 | 2015-10-17 05:49:00 |
| 0011 | 4 | 2015-10-17 04:38:00 | 2015-10-17 04:39:00 |
| 0006 | 3 | 2015-10-17 03:07:38 | - |
| 000C | 3 | 2015-10-17 02:02:56 | - |
| 0009 | 1 | 2015-10-17 01:10:27 | - |
| 000F | 6 | 2015-10-17 00:32:06 | - |
| 0005 | 2 | 2015-10-16 23:47:16 | 2015-10-16 23:48:16 |
| 0014 | 4 | 2015-10-16 23:46:03 | 2015-10-16 23:48:03 |
| 0019 | 3 | 2015-10-16 23:34:20 | 2015-10-16 23:34:50 |
| 0019 | 1 | 2015-10-16 23:34:19 | 2015-10-16 23:34:49 |
| 000F | 5 | 2015-10-16 23:30:06 | 2015-10-16 23:32:06 |
| 000F | 1 | 2015-10-16 23:30:04 | 2015-10-16 23:32:04 |
| 0014 | 3 | 2015-10-16 21:05:30 | 2015-10-16 21:10:30 |
| 0007 | 1 | 2015-10-16 18:47:45 | 2015-10-16 19:47:45 |
| 0007 | 4 | 2015-10-16 18:47:45 | 2015-10-16 19:47:45 |
| 0014 | 5 | 2015-10-16 17:26:13 | 2015-10-16 17:26:43 |
| 0016 | 1 | 2015-10-16 17:21:28 | 2015-10-16 17:23:28 |
| 000C | 4 | 2015-10-16 17:12:18 | 2015-10-16 17:12:48 |
| 000C | 2 | 2015-10-16 17:12:17 | 2015-10-16 17:17:17 |
| 0007 | 3 | 2015-10-16 16:17:19 | - |
| 0011 | 2 | 2015-10-16 15:40:15 | 2015-10-16 15:42:15 |
| 0018 | 4 | 2015-10-16 14:21:24 | - |
| 0007 | 2 | 2015-10-16 12:36:25 | - |
| 000D | 1 | 2015-10-16 12:33:41 | 2015-10-16 13:33:41 |
| 000E | 3 | 2015-10-16 11:40:07 | 2015-10-16 13:40:07 |
| 000D | 3 | 2015-10-16 09:55:55 | - |
| 000C | 1 | 2015-10-16 09:28:08 | - |
| 0003 | 1 | 2015-10-16 09:19:50 | 2015-10-16 10:19:50 |
| 0006 | 1 | 2015-10-16 08:34:04 | - |
| 0017 | 2 | 2015-10-16 07:57:38 | - |
| 000F | 4 | 2015-10-16 07:56:27 | - |
| 0018 | 3 | 2015-10-16 07:52:37 | 2015-10-16 07:54:37 |
| 0018 | 1 | 2015-10-16 07:52:36 | 2015-10-16 07:53:36 |
| 000E | 2 | 2015-10-16 06:23:18 | 2015-10-16 06:28:18 |
| 0019 | 2 | 2015-10-16 05:01:43 | - |
| 0013 | 3 | 2015-10-16 04:21:11 | - |
| 0013 | 2 | 2015-10-16 03:28:17 | - |
| 0013 | 4 | 2015-10-16 03:10:35 | 2015-10-16 03:11:35 |
| 0014 | 2 | 2015-10-16 03:04:01 | 2015-10-16 03:09:01 |
| 000F | 3 | 2015-10-16 01:42:43 | - |
| 0011 | 1 | 2015-10-16 01:20:02 | - |
| 000E | 4 | 2015-10-15 23:58:14 | 2015-10-16 00:03:14 |
| 000E | 1 | 2015-10-15 22:50:11 | - |
| 0011 | 5 | 2015-10-15 | 2015-10-15 |

|  |  | 19:15:10 | 19:16:10 |
|---|---|---|---|
| 0005 | 1 | 2015-10-15 17:46:07 | 2015-10-15 18:46:07 |
| 0005 | 4 | 2015-10-15 13:00:01 | 2015-10-15 5:00:01 |

**Table 2. Node information with location and address [16]**

| Name | Location | MAC Address | Short Address |
|---|---|---|---|
| 0002 | Blk-B | 0x0002000200020002 | 0x0002 |
| 0003 | North Wing | 0x0003000300030003 | 0x0003 |
| 0004 | Corridor-B | 0x0004000400040004 | 0x0004 |
| 0005 | B-A | 0x0005000500050005 | 0x0005 |
| 0006 | South Wing | 0x0006000600060006 | 0x0006 |
| 0007 | Blk-A | 0x0007000700070007 | 0x0007 |
| 0008 | Car Park | 0x0008000800080008 | 0x0008 |
| 0009 | Nw Lab | 0x0009000900090009 | 0x0009 |
| 000A | Corridor-A | 0x000A000A000A000A | 0x000A |
| 000B | South Wing | 0x000B000B000B000B | 0x000B |
| 000C | South Wing | 0x000C000C000C000C | 0x000C |
| 000D | Blk-B | 0x000D000D000D000D | 0x000D |
| 000E | South Wing | 0x000E000E000E000E | 0x000E |
| 000F | Blk-B | 0x000F000F000F000F | 0x000F |
| 0010 | North Wing | 0x0010001000100010 | 0x0010 |
| 0011 | North Wing | 0x0011001100110011 | 0x0011 |
| 0012 | Corridor-B | 0x0012001200120012 | 0x0012 |
| 0013 | Blk-A | 0x0013001300130013 | 0x0013 |
| 0014 | Nw Lab | 0x0014001400140014 | 0x0014 |
| 0015 | Nw Lab | 0x0015001500150015 | 0x0015 |
| 0016 | Corridor-A | 0x0016001600160016 | 0x0016 |
| 0017 | ECE Dept | 0x0017001700170017 | 0x0017 |
| 0018 | Nw Lab | 0x0018001800180018 | 0x0018 |
| 0019 | Corridor-B | 0x0019001900190019 | 0x0019 |